# Adsorption of Br$_2$ onto Small Au Nanoclusters

C. R. Salvo, J. Keagy and J. A. Yarmoff*

*Department of Physics and Astronomy, University of California, Riverside, CA 92521*

## ABSTRACT

Au nanoclusters grown on SiO$_2$ by physical vapor deposition are exposed to Br$_2$ and then measured with 1.5 keV Na$^+$ low energy ion scattering. It is found that the clusters are able to dissociate the molecules which then adsorb as individual Br atoms, but Br$_2$ does not stick to the bare substrate nor to bulk Au. Adsorption is the first step in any surface chemical reaction, and this result shows how nanoclusters can induce adsorption of species that otherwise do not stick. Results from the literature indicate that catalysis involving nanoclusters occurs at the edges and that the edge atoms are positively charged. This information in conjunction with the ion scattering results lead to the conclusion that the Br adatoms are negatively charged and ionically bonded at the edges of the clusters. Br$_2$ is also a known catalytic poison and this work shows how its adsorption blocks sites that would otherwise be involved in nanocatalysis.

*Corresponding author, E-mail: yarmoff@ucr.edu

1. **INTRODUCTION**

The adsorption of a molecule onto a solid surface is a fundamental, and usually the first, step in many chemical and physical processes, particularly in applications such as catalysis,[2] etching,[3-4] and chemical vapor deposition.[5] Here, the dissociative adsorption of $Br_2$ is shown to occur strongly in the presence of small Au nanoclusters supported on a silicon dioxide ($SiO_2$) substrate, but not with bulk Au metal nor on the bare substrate. Small metal nanoclusters supported on oxides are extremely effective nanocatalysts, with rates that rival those of enzymes in biological systems.[6] This result thus provides a glimpse into how metal nanoclusters promote the adsorption of precursor species at the beginning of the chemical reaction process. In addition, $Br_2$ is known to poison catalysis employing nanoclusters,[7] so that an understanding of its adsorption also reveals information about the inner workings of the nanocatalytic mechanism.

The surfaces are investigated primarily with alkali low energy ion scattering (LEIS) in two different modes. First, LEIS spectra are used to reveal the distribution of elements at the surface.[8] Second, a novel application of LEIS in which the neutralization of scattered alkali ions is measured provides information on the surface electronic properties of the nanoclusters. The neutralization is sensitive to the surface local electrostatic potential (LEP) a few Å's above the target atom;[9-10] this technique has been applied to scattering from clean metals,[11-13] adsorbates on metals and semiconductors,[9, 14-15] and metal nanoclusters.[16-21]

For nanoclusters deposited on oxide substrates, the neutralization probability of scattered alkali ions is particularly sensitive to the size of the clusters in that it is high for the smallest clusters and decreases as the clusters grow larger.[16-19] In recent work, we showed that the high neutralization is due to the fact that the low coordinated edge atoms of the nanoclusters are positively charged, while the center atoms are nearly neutral.[22] The positive charge creates upward pointing dipoles that decrease the local electrostatic potential (LEP) above the edge atoms causing a higher neutralization probability for alkali ions scattered from those atoms. In contrast, the ions that scatter from center atoms have a low neutralization probability due to the high work function of neutral Au. Since the ratio of the number of edge atoms to center atoms decreases with cluster size, the overall neutralization probability decreases with cluster size. This idea was confirmed by a simple calculation of the neutral fraction as a function of cluster size that quantitatively matches experimental data. Since it has been shown by others that the edge atoms are the active sites for chemisorption during catalytic reactions involving nanoclusters,[23] this result suggests that there is a relationship between the charge state of the edge atoms and their ability to promote nanocatalysis.

The adsorption of $Br_2$ molecules onto a solid is a dissociative process that involves the scission of the Br-Br bond and thus requires a reactive surface.[24] The fact that dissociative adsorption readily occurs on small Au nanoclusters, while it does not occur on the bare substrate nor readily on the bulk metal surface, indicates that the nanoclusters are directly involved in the Br-Br bond cleavage, and this chemical reactivity is thus related to the catalytic behavior of small Au nanoclusters. A reduction in the neutralization of scattered $Na^+$ with $Br_2$ exposure shows that the average LEP above the Au edge atoms decreases in the presence of adsorbed Br. This leads to the conclusion that $Br_2$ dissociatively chemisorbs by forming ionic bonds to the positively charged edge atoms. Bonding to the edge atoms produces downward pointing dipoles that oppose those formed by the positively charged edge atoms and reduces the neutralization of ions scattered from those sites. The fact that Br forms such bonds suggests that the charge state of the atoms in a nanocatalyst plays an important role in the adsorption step of surface chemical reactions involving nanoclusters.

## 2. EXPERIMENTAL PROCEDURE

The experiments are conducted in an ultra-high vacuum (UHV) chamber that has a base pressure less than $1 \times 10^{-9}$ Torr. The Si(111) substrate is mounted on the foot of an XYZ rotary manipulator, and is electrically isolated so that current can be run directly through the wafer for resistive heating. There is a load-lock chamber and sample transfer system attached to the chamber so that new samples can be quickly introduced for each measurement without breaking vacuum in the main sample preparation and analysis chamber. The main chamber includes equipment for performing time-of-flight (TOF) LEIS, low energy electron diffraction (LEED) and x-ray photoelectron spectroscopy (XPS).

The substrate is a 5x5x1 mm$^3$ single crystal Si(111) wafer (*n*-type, 5-10 Ω cm) onto which a SiO$_2$ film is grown *in situ*. After insertion into the UHV chamber, the Si samples are initially degassed by running 0.5 A through them (roughly 250°C) for a minimum of 30 min to remove adsorbed water and hydrocarbons. The native oxide layer and any more strongly bound contaminants are then removed by "flashing" the sample with 9 A. This value is chosen by systematically increasing the current and monitoring the O 1s peak with XPS after flashing until there is no longer any oxygen signal present. The surfaces cleaned in this manner show no contamination with XPS and display clear 7x7 LEED patterns.[25] The oxide layer is then grown by heating the sample to approximately 700°C under $2 \times 10^{-5}$ Torr of flowing O$_2$ for 30 min, which produces a uniform thermal oxide layer.[26] Note that the samples are cooled with the O$_2$ still present to avoid the formation of pinholes in the films.[27] XPS is used to confirm the growth of a SiO$_2$ layer and provide its thickness.

Au is deposited onto the sample via evaporation from a heated tungsten filament (Mathis) with Au wire (99.99%) wrapped around it. The Au coverage is calculated by calibrating the deposition rate using a quartz crystal microbalance (QCM) with the assumption that 1 monolayer (ML) of Au corresponds to a single atomic layer of Au(111) with a density of 19.3 g/cm$^3$, which has a height of 2.6 Å.[28] The reported amounts of deposited Au correspond to what the thickness of a Au film would be if it grew in a layer-by-layer mode. Since it actually forms nanoclusters, however, the coverage values are useful as a guide to the amount of Au that is deposited, but are not directly related to the thickness of the clusters.

Br$_2$ molecules are produced from a solid-state electrochemical cell based on a AgBr pellet affixed to Ag foil.[29-30] The exposures are given in units of μA-min, which refers to the integrated current run through the cell. It had been previously reported that a 10 μA-min exposure corresponds approximately to 1 molecule impacting each surface atom,[31] but that estimate is dependent on the specific cell parameters and geometry. No detailed calibration of the exposure was possible in the present setup as the sticking coefficient of Br$_2$ is unknown.

XPS measurements are made using a Riber Mg K$_α$ (1253 eV) x-ray source that has a natural line width of 1.10 eV. A Riber cylindrical mirror analyzer (CMA) with adjustable resolution is used to measure the energy distribution of the emitted photoelectrons. The XPS spectra presented here are collected with an energy resolution of 3.0 eV, although a resolution that matches the x-ray source linewidth is employed to collect detailed spectra that are used to measure the film thickness.

TOF-LEIS is performed with the sample held at room temperature using 1.5 keV $^{23}$Na$^+$ alkali ions, similar to previous descriptions.[9, 19] The Na$^+$ ions are generated from a thermionic emission source (Kimball Physics) and incident at 30° degrees to the surface normal. The beam is pulsed at 80 kHz by deflecting it across a 1 mm diameter aperture mounted in front of the gun. The particles emitted along the surface normal, which are scattered at a 150° angle, are collected by a triple microchannel plate (MCP) array located at the end of the 0.43 m long flight leg. Deflection plates mounted in the leg are used to distinguish between scattered neutral and charged particles. With both plates grounded, all of the particles pass through to the detector, but only the neutral particles pass when 400 V is placed across the plates. "Total" and "neutral" yield TOF spectra are collected by switching the deflection plate voltage on and off every 60 s during the approximately 15 min it takes to collect a spectrum, which avoids any effects of long-term drift in the ion beam current. The entrance to the MCP is held at ground potential to ensure equal sensitivity to charged and neutral particles, but there is a decrease in absolute sensitivity as the kinetic energy of the scattered projectiles falls below about 1 keV.[32]

## 3. RESULTS

Figure 1 shows XPS spectra used to confirm the sample cleanliness and monitor the presence of Au and Br. Spectrum (a) was collected after exposing clean Si(111) to 60 μA min of $Br_2$, and a clear Br 3d peak is visible. This is expected since Br readily adsorbs on clean Si.[24, 33] Spectrum (b) was collected from an $SiO_2$ film grown on Si(111), which confirms the cleanliness of the sample and enables a measurement of the film thickness. The thickness is determined from the ratio of the $SiO_2$ to the bulk Si 2p component in a high-resolution Si 2p spectrum (not shown) to be approximately 0.7 nm using 2.5 nm for the photoelectron escape depth.[26] Spectrum (c) was collected after 0.30 ML of Au was deposited onto the $SiO_2$, which is a coverage that forms Au nanoclusters,[34] and it shows no features other than those indicative of Si, O and Au. The final spectrum (d) was collected after exposure of the Au nanocluster covered-surface to 40 μA min of $Br_2$. Note that there is a small feature near the position of the Br 3d peak, but it is also present in the spectrum collected before $Br_2$ exposure and is likely a satellite feature associated with the Au 4f level as is common in XPS due to a small amount of higher energy x-rays emitted by the Mg K$_\alpha$ source. Note that any Br 3d signal would be negligible because there is only a submonolayer coverage of Br and XPS is not sensitive enough to reveal such a small amount.

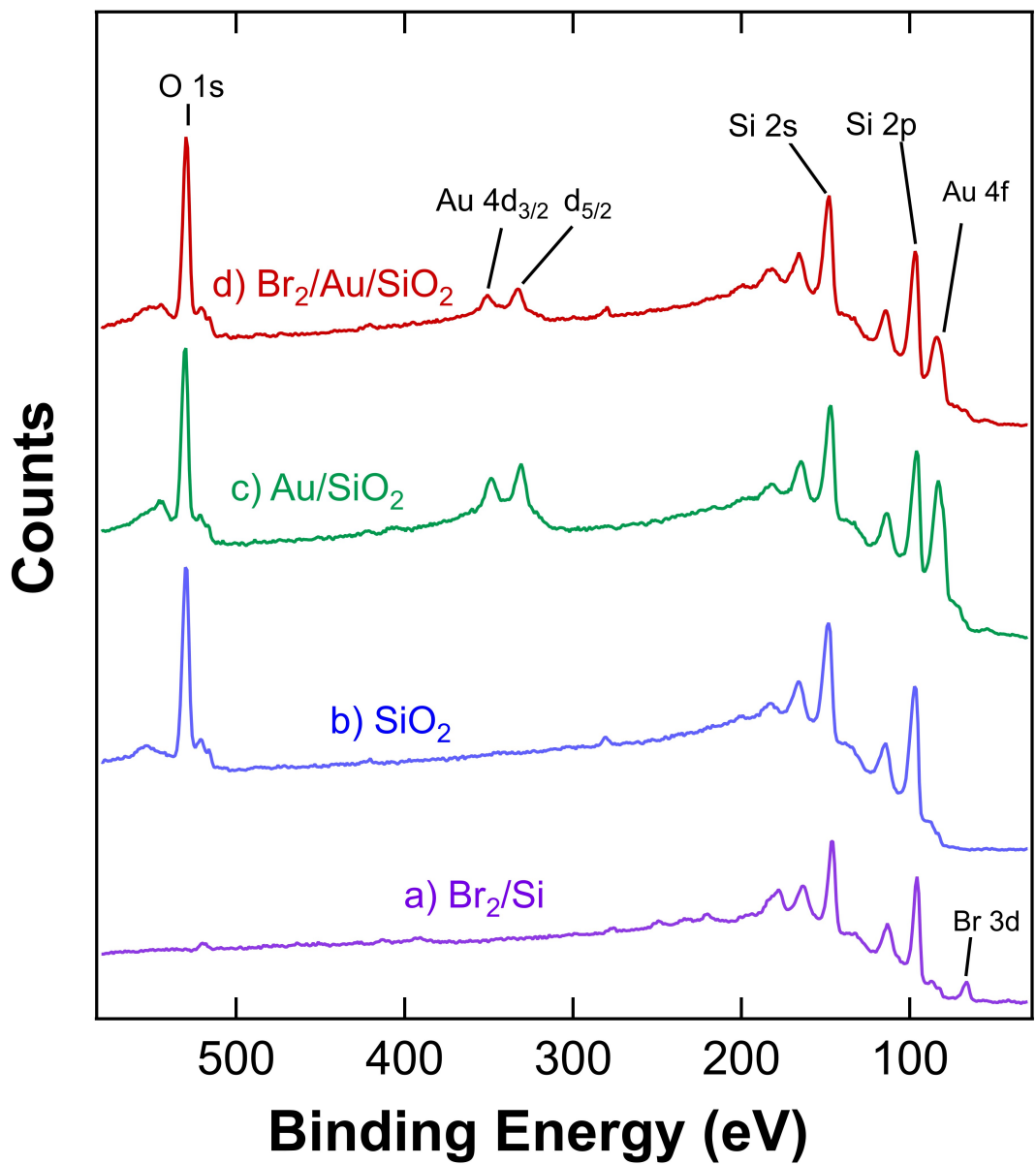

**Figure 1.** Typical XPS spectra collected from (a) Si(111) exposed to 40 µA min of Br$_2$, (b) thermally prepared SiO$_2$ on Si(111), (c) 0.30 ML Au deposited onto SiO$_2$, and (d) that surface exposed to 60 µA min of Br$_2$. The primary core levels associated with Si, O, Au and Br are indicated in the figure.

LEIS is an extremely surface sensitive technique,[12] even more so than XPS. This is largely due to shadowing and blocking,[35-36] which effectively only allow backscattered projectiles to be singly scattered from the first few layers, although multiply scattered projectiles can probe deeper into the material. The role of shadowing and blocking for a particular surface depends on the geometry of the setup and the crystal structure, so that the depth sensitivity is system dependent. It was previously shown that for $Na^+$ scattered from Au nanoclusters on an oxide substrate, single scattering probes only the outermost Au atoms in each cluster.[22]

Figure 2 shows representative TOF-LEIS spectra of the total scattered yield of Na projectiles collected from three different samples. The x-axis was converted from time to energy using the known length of the flight tube. The spectra show single scattering peaks (SSPs), which provide a measure of the surface composition. Single scattering occurs when a projectile undergoes a hard binary collision with a surface target atom and scatters directly into the detector, and is the primary means for determining surface composition via LEIS.[12, 37] The energy of the scattered projectile can be calculated classically considering only conservation of energy and momentum in an elastic collision with an unbound target atom, which predicts that Na will scatter from the heavier Au atoms with a higher energy than when scattered from Br. Note that there is also some energy lost to inelastic processes, but the amount is small compared to the elastic energy loss.[38]

The bottom spectrum in Fig. 2 was collected following exposure of clean $SiO_2$ to $Br_2$. The absence of Si and O SSPs is expected because their masses are too small to produce singly scattered Na at a 150° angle with a large enough energy to be detected. Na scattered from Br does have enough energy to be detected, however, so that the absence of a Br SSP in the spectrum indicates that $Br_2$ does not stick to $SiO_2$. XPS spectra collected from this sample (not shown) also had no discernable Br signal. The inertness of $SiO_2$ to $Br_2$ adsorption is presumably because the amorphous oxide surface does not have reactive sites that can break the halogen-halogen molecular bond.

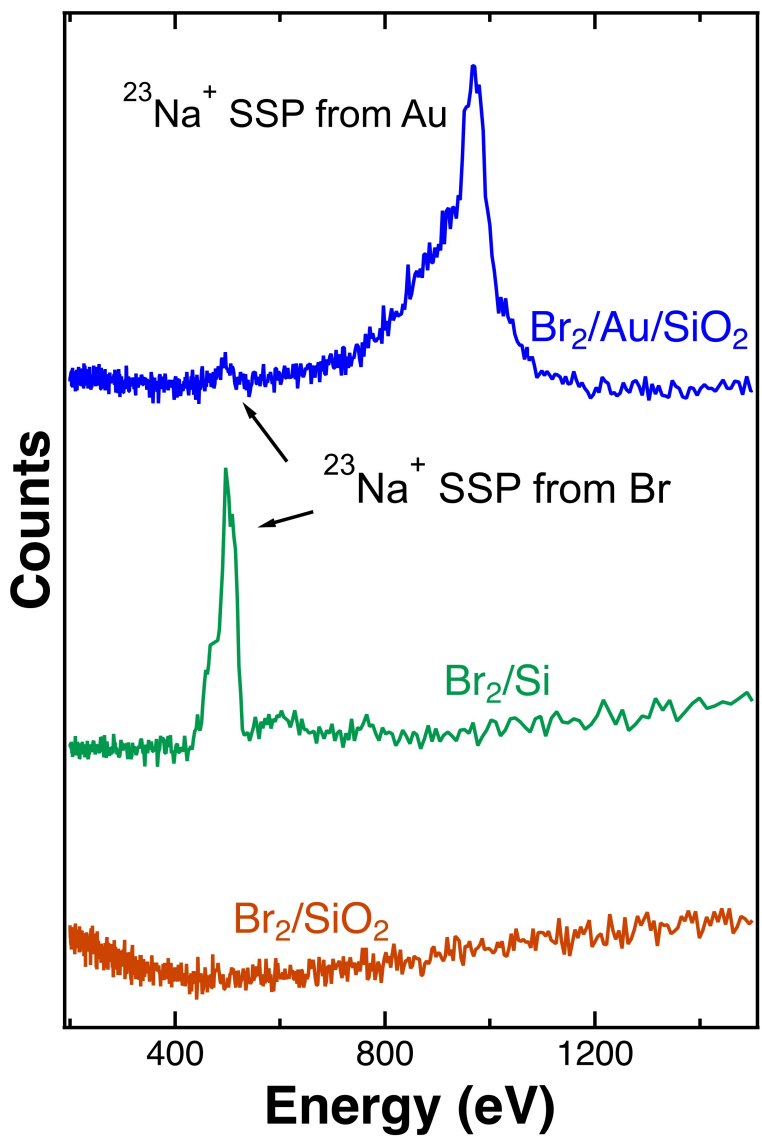

**Figure 2.** Typical TOF LEIS spectra collected using 1.5 keV Na$^+$ projectiles. The Au and Br SSPs are indicated when present. The three spectra were collected after a 30 μA min Br$_2$ exposure of a SiO$_2$ film, clean Si(111)-7x7, and nanoclusters produced by deposition of 0.53 ML of Au on SiO$_2$.

The middle spectrum in Fig. 2 was collected from Si(111)-7x7 exposed to 60 μA min of Br$_2$. The SSP for Br is clearly seen at 496 eV, indicating that Br$_2$ readily adsorbs on clean Si(111), consistent with the XPS results above and reports in the literature.[24, 33, 39] The dangling bonds on the Si(111)-7x7 surface provide the reactive sites that break the Br-Br bonds and lead to dissociative chemisorption.

The upper spectrum in Fig. 2 was collected from Au nanoclusters on $SiO_2$ following $Br_2$ exposure. The sample was prepared by deposition of 0.19 ML of Au onto $SiO_2$, which is a coverage that produces active nanoclusters, and then exposing it to 40 µA min of $Br_2$. The strongest peak in the spectrum is at 968 eV, which is the Au SSP. Although the Br SSP signal is much smaller than for bromine-exposed Si, it is clearly visible. Note that the MCP sensitivity and differential cross section must be considered when comparing the relative sizes of different SSPs to determine the absolute amount of adsorbed Br, as quantified below.

In addition, a polycrystalline Au foil was cleaned by $Ar^+$ ion sputtering and then exposed to $Br_2$. This foil shows no evidence of Br with XPS and a very small Br SSP in LEIS (not shown) that indicates the adsorption of approximately 1 Br adatom per 200 Au surface atoms (the method used for this calculation is described below). This implies that polycrystalline Au is essentially inert to reaction with $Br_2$. Although there are reports of $Br_2$ adsorption onto clean Au(100) in vacuum, which has an unusual 5x20 surface unit cell,[40-41] there are, to our knowledge, no reports of $Br_2$ adsorption on other single crystal faces of Au nor on polycrystalline Au. There are, however, reports of $Cl_2$ and $I_2$ chemisorption on Au(111).[42-43]

The intensities of the SSPs in the total yield spectra are used to determine the surface coverages of Br and Au. Each SSP is integrated following subtraction of the multiple scattering background, as described elsewhere.[9] The uncertainty in each SSP area is assumed to be purely statistical so that the error bars are set to the square root of all the counts, including the background. When comparing the Br and Au SSPs, corrections also need to be made to account for their relative sensitivities. First, the MCP sensitivity depends on the kinetic energy of the scattered projectile, which depends, in turn, on the mass of the target atom. The values of the MCP efficiency at the relevant kinetic energies are 0.30±0.01 and 0.50±0.01 for the Br and Au SSPs, respectively.[32] Second, Na scattered from Br and Au have different scattering cross sections, which are determined using Thomas-Fermi theory.[44] Such a calculation shows that the differential cross section for 1.5 keV Na scattered at 150° from Au is 2.15 times larger than for scattering from Br. Thus, when reporting the ratio of Br to Au, the raw ratios of the integrated SSP areas are multiplied by 3.58 to compensate for the differences in the MCP sensitivities and differential cross sections.

It is well established that Au atoms deposited onto $SiO_2$ follow a Volmer-Weber growth mode and form nanoclusters rather than a dispersed film.[19, 34] The growth of nanoclusters by direct deposition on an oxide surface produces a narrow range of cluster sizes, and the average size increases as more Au is deposited.[1, 34] The amount of Au deposited is converted to the average diameter of the nanoclusters using STM data from the literature for Au deposited on $TiO_2$.[1]

Figure 3 plots the ratio of Br to Au atoms at the outermost surface of the nanoclusters as a function of the diameter of the clusters following exposures to 30 and 40 μA min of $Br_2$. The ratio is generated by dividing the integrated Br and Au SSP's from the same TOF total yield spectrum and correcting for the MCP sensitivities and scattering cross sections. The error bars are calculated by propagating the statistical error determined for the area of each individual SSP.

The inset in Fig. 3 shows the ratio of surface Br to Au as a function of $Br_2$ exposure for a fixed Au coverage of 0.30 ML, which corresponds to a 3.0 nm average cluster diameter. The data initially increases linearly showing that the sticking coefficient is constant until Br occupies all of the adsorption sites. At this point, no more $Br_2$ sticks and the surface coverage saturates. For 0.30 ML of Au, this occurs at a $Br_2$ exposure around 50 μA min. Although the exposure needed to reach saturation is likely to depend on the nanocluster size, a 50 uA min exposure is assumed to be sufficient to attain saturation for the range of cluster sizes studied here.

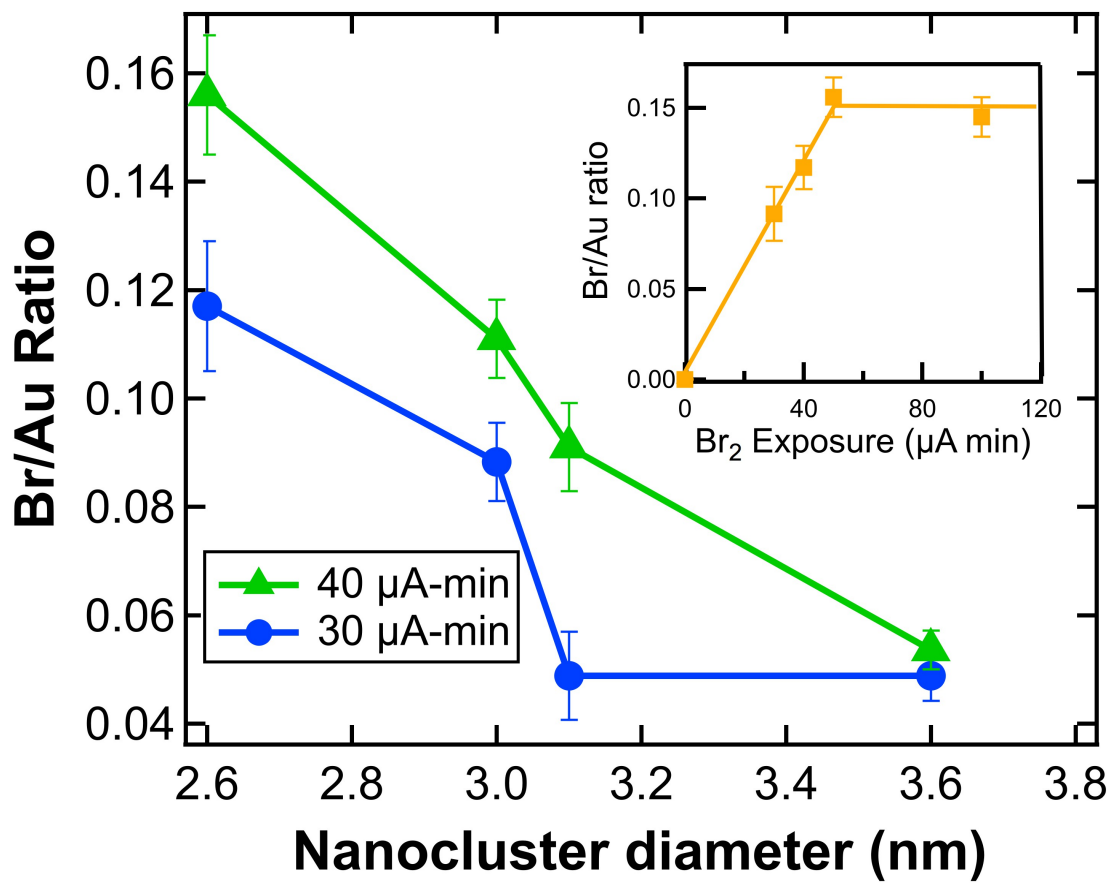

**Figure 3.** The ratio of the number of outermost Br to Au atoms, determined from the ratio of the total yield Br and Au SSPs for scattered 1.5 keV Na$^+$ after normalization by the sensitivity of the MCP and the scattering cross sections, shown as a function of the average Au nanocluster diameter on SiO$_2$. Inset: The ratio of Br to Au shown as a function of Br$_2$ exposure following deposition of 0.30 ML of Au on SiO$_2$, which forms 3.0 nm diameter clusters.

4.  DISCUSSION

With the use of the data in Fig. 3, estimates of the number of Br atoms per nanocluster and per edge atom are made and provided in Table 1. The accuracy of these calculations is limited by a few factors. For example, there may be a systematic error due to differences in the cluster formation process on the SiO$_2$ substrate used in the present experiment as opposed to the TiO$_2$ substrate used to collect the STM images of Lai et al.,[1] from which the cluster sizes are calculated here. Although the two substrates are different, it was shown that the average cluster size formed on SiO$_2$ and TiO$_2$ are very similar for 0.20 ML of deposited Au,[45] however, so that any systematic error is likely to be small. In addition, the accuracy of the QCM calibration of the deposition rate can lead to another small systematic error.

**Table 1.** The Au surface coverages, Br$_2$ exposures, cluster sizes, number of outermost Au and edge atoms per cluster and coverages of Br determined from the LEIS data. The cluster sizes and number of edge atoms are determined from the amount of Au deposited using STM images from the literature as a calibration (see text).[1] The average numbers of outermost Au atoms per cluster are calculated from the sizes of the Au SSPs. The ratios of Br to Au, as shown in Fig. 3, are used to determine the average numbers of Br atoms per cluster and the Br surface coverages.

| **Au deposited (ML)** | 0.19 | 0.30 | 0.53 | 0.79 | 1.10 |
|---|---|---|---|---|---|
| **Br exposure (µA-min)** | 40 | 40 | 40 | 40 | 60 |
| **Average nanocluster diameter (nm)** | 2.6±0.2 | 3.0±0.2 | 3.1±0.2 | 3.6±0.2 | 3.7±0.2 |
| **Average outermost Au atoms per cluster** | 58.9±9.5 | 75.1±14.2 | 81.0±14.0 | 113±13.0 | 119±13.0 |
| **Average Br atoms per cluster** | 9.2±1.6 | 10.4±1.4 | 7.3±0.7 | 7.3±1.0 | 3.4±0.6 |
| **Average edge Au atoms** | 27.2±2.1 | 33.5±2.1 | 34.6±2.1 | 39.8±2.1 | 42.0±2.1 |
| **Average Au edge atom per Br atom** | 3.0±0.9 | 3.2±0.9 | 4.7±0.8 | 5.5±0.9 | 12±0.9 |
| **Au surface coverage** | 13±2% | 20±2% | 24±3% | 40±4% | 43±4% |
| **Br surface coverage** | 2.0±0.2% | 2.8±0.2% | 2.4±0.3% | 2.6±0.4% | 1.2±0.1% |

Table 1 shows that the number of Br adatoms is much smaller than the number of Au atoms at the surfaces of the nanoclusters. This implies that Br does not completely cover the nanoclusters in the same way that it covers a Si(111) surface exposed to Br$_2$.[24] One explanation for the very small amount of Br adsorption is that the atoms only attach at the edges of the clusters, consistent with the current consensus that adsorption in nanocatalytic reactions occurs at the edges.[46-50]

To determine the neutralization probability, i.e. the neutral fraction (NF), of $Na^+$ singly scattered from Au, the integrated SSP of the neutral yield spectrum is divided by that of the total yield spectrum. The neutralization probability depends on the LEP at a particular distance that the projectile is above the target atom along the exit trajectory, the ionization level of the projectile, and the kinetic energy and exit angle of the scattered particle.[11, 14] When an alkali projectile approaches a surface, its ionization level shifts up as it sees its image charge in the surface. At the same time, the sharp *s* projectile ionization level begins to hybridize with orbitals in the surface causing it to broaden.[13, 51] The overlap between the broadened *s* level and the filled states in the surface allow electrons to tunnel back and forth once the projectile is close enough to the surface. Because the velocity of the projectile is large compared to the electron-tunneling rate, the interaction occurs non-adiabatically and the charge distribution of the scattered projectiles is determined along the exit trajectory at a "freezing point" that is typically a few Å above the surface.[10, 52] This distribution is measured as the NF of the scattered alkali projectiles, which increases as the LEP decreases, and vice versa.

When scattering from a homogeneous surface, such as a clean metal, the LEP is the same everywhere so that alkali projectiles scattered from any surface site will have the same NF, and it can be calculated using the global surface work function.[53] If the surface LEP is inhomogeneous, however, then the NF for a particular scattering event depends on the LEP above the target atom, which is sometimes referred to as the local work function. This point is clearly illustrated in TOF-LEIS measurements from metal surfaces with alkali adsorbates.[9] When an alkali atom adsorbs onto a metal, it donates its valence electron to the substrate and creates a local upward pointing dipole at the adatom causing the LEP above that site to decrease from that of the surrounding surface. When using alkali LEIS, it is possible to explicitly differentiate the NF of the adsorbate and substrate by their scattered energies.[9, 54] It was found that alkali ions scattered from alkali adatoms have a larger NF than for scattering from the substrate when the coverage is small, consistent with the decrease in LEP associated with the positively charged adatoms.

A recent report from our group[22] agrees with work from the literature in that the clusters contain positively charged Au atoms,[55-59] although others have concluded that the clusters are overall negatively charged.[60-63] It was calculated from density functional theory (DFT) that the edge atoms of a nanocluster are the ones that are positively charged[59, 64] and upward pointing dipoles are thus created at those sites. These upward pointing dipoles create a lower LEP, as with alkali adatoms, which raises the NF of the projectiles scattered from the edge atoms. In contrast, the center atoms are nearly neutral so that alkali ions scattered from them have the low NF associated with scattering from a bulk Au metal surface. The average NF obtained from a LEIS spectrum is thus larger in scattering from nanoclusters than for scattering from bulk Au due to the reduced LEP above the edge atoms. Because the ratio of edge to center atoms decreases with cluster size, the NF decreases with size. The observation of a reduction in the NF of scattered alkali ions with cluster size thus provides an indication that it is the edge atoms that are positively charged.

Figure 4 shows the NF as a function of $Br_2$ exposure for different Au nanocluster sizes. As expected, in the absence of $Br_2$ exposure the NF is highest for the smallest clusters, and it reduces as the cluster size increases with further Au deposition.[17, 19] For all of the Au cluster sizes, the NFs in Fig. 4 decrease significantly with $Br_2$ exposure, indicating that Br adsorbs even on the smallest nanoclusters. In addition, the NF does not change beyond the limits set by the error bars after a $Br_2$ exposure of 40 µA min or more, indicating that such an exposure produces an amount of adsorbed Br that is close to saturation coverage.

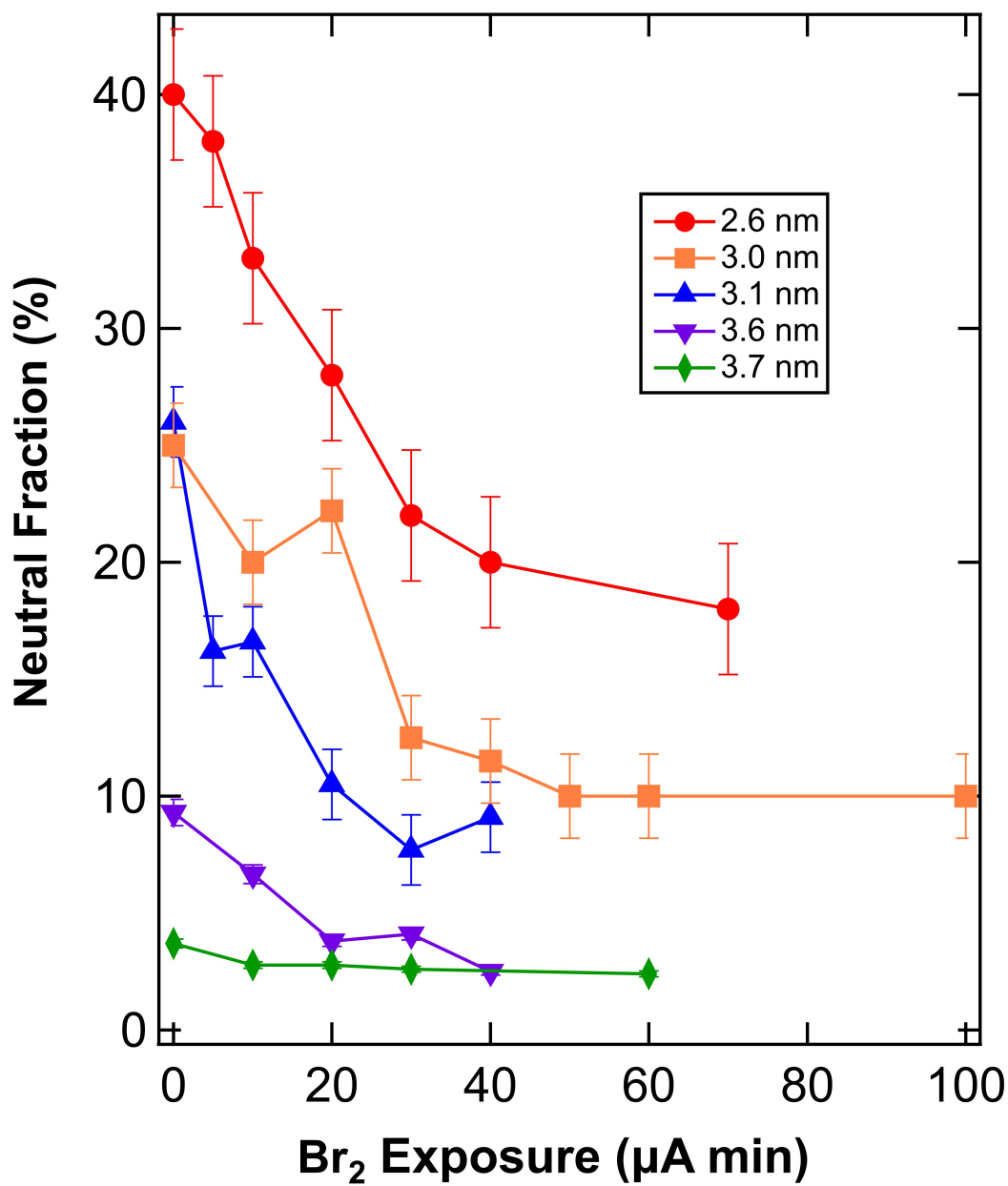

**Figure 4.** The neutral fractions of 1.5 keV Na$^+$ projectiles singly scattered from Au atoms in nanoclusters on SiO$_2$ with the indicated average diameters, shown as a function of Br$_2$ exposure.

The data in Table 1 show that the total number of adsorbed Br atoms per cluster is relatively constant for clusters with average diameters in the range of 2.6 to 3.6 nm. Lemire *et al.* found that for 2 nm diameter Au nanoclusters on FeO(111), 5 CO molecules adsorb per nanocluster and do so at the low coordinated Au edge atom sites.[50] As calculated in Table 1, between 7 and 10 Br atoms adsorb per nanocluster for 0.19 to 0.79 ML of deposited Au. Since the number of CO molecules is comparable to that of Br atoms per nanocluster, it is surmised that the Br atoms occupy the same edge sites as adsorbed CO. A drop off in the dipole strength of the nanocluster edge atoms occurs with 1.10 ML of Au due to inter-nanocluster effects that occur when the clusters are close to each other, as discussed previously,[22] thereby reducing their ability to dissociate $Br_2$ and thus decreasing the amount of Br that adsorbs per cluster.

It has been shown that the nanoclusters' catalytic activity is a function of cluster size with a maximum at a particular diameter.[47, 49] For example, Choudhary *et al.* found that Au nanoclusters on $TiO_2$ with an average diameter of 3.2 nm on $TiO_2$ produce the highest turnover frequency for the CO oxidation reaction.[65] Many ideas have been proposed to explain the catalytic activity of metal nanoclusters, but there is not a complete consensus among researchers that would accompany a full understanding of the underlying mechanisms.[66-68] For example, the charge state of the supported nanoclusters and the involvement of that charge in surface reactions are still under debate.[47, 49]

The first step in a catalytic reaction is the adsorption of a precursor molecule onto a surface, such as CO adsorbing onto Au nanoclusters in the oxidation reaction, which has been extensively studied.[50, 65, 69-70] Since it is likely that Br adatoms occupy the same sites when adsorbed on a nanocluster, it can be concluded that halogens poison nanocatalysts by blocking the active sites. Furthermore, it can be inferred that the mechanisms responsible for the adsorption of the precursors and the poisons are similar and thus have comparable saturation coverages.[71]

Poisoning can occur by a physical blockage preventing adsorption of the precursor molecule which can entail electronic modification of the nearest neighbor atoms by that adsorbate, restructuring of the adsorbent surface, or hindering of surface diffusion of the adsorbed reactants.[72] Zhu *et al.* tested the reactivity of Pt nanoclusters on $TiO_2$ by oxidizing formaldehyde both before and after exposure to halogens.[7] From the testing of formaldehyde oxidation using Pt nanoclusters as catalysts, they found that any of the four halogens causes the performance of the nanocatalyst to worsen. In addition Gracia *et al.* demonstrated the deactivation of Pt nanoclusters used for CO oxidation by Cl adsorption.[73]

Br and CO have some similarities when adsorbing onto Au nanoclusters, as discussed above, making it reasonable to conclude that the Br adsorbs at the edges of the nanoclusters. Once $Br_2$ is near enough to the edge of a nanocluster, the positively charged edge atoms can lead to a configuration of adsorbed Br atoms that is lower in energy than the $Br_2$ molecule, causing it to dissociate. This is analogous to the dissociative chemisorption of halogens on a flat solid surface, where charge is donated locally from the bonding atom to the halogen adatom to form a partially ionic bond.[74] When a Br atom is close to the substrate, it can pick up an electron to form a filled shell negative ion with a charge of -1.0 $e$. Since the average Bader charge of the edge atoms in a Au nanocluster on $TiO_2$ is calculated to be about +0.4 $e$,[59] it is possible that the Br ionically bonds in a bridge site between two edge atoms as the magnitude of the charge of two Au edge atoms is nearly equal to the magnitude of a single Br negative ion.

If the Br adatoms were to adsorb ionically at the edge sites, the reduction in the NF can be explained in terms of the LEP change induced by the adatoms. An additional downward dipole would be formed by the negative Br adatom and the positively charged Au edge atom, which would increase the LEP above the edge atoms from that of the bare clusters and therefore decrease the neutralization probability of $Na^+$ when scattering from those edge atoms. Since the NF for the smaller clusters does not fully decrease to that of Au metal, the saturation Br coverage is not sufficient to cover every edge atom, consistent with the data in Table 1.

If the saturation coverage were determined solely by the availability of positive sites along the edge of the nanocluster, it would imply that the saturation coverage should correspond to one Br atom per two Au edge atoms. The actual ratios lie between one Br per 3.0 to 5.5 Au edge atoms, however. This could be partially due to the fact that the 40 µA min exposures used for some of the data in Table 1 are below what is needed to obtain full saturation of Br. Because these exposures are so close to saturation, however, it is more likely that there is an intrinsic limit to the coverage due to other effects. For example, an increase in energy due to repulsion between neighboring negatively charged Br adatoms could lead to a saturation coverage that is less than one Br atom per two Au edge atoms.

## 5. CONCLUSIONS

It is found that $Br_2$ readily adsorbs to small Au nanoclusters, implying that the Br-Br bond breaks because of the catalytic ability of the clusters. It is concluded that the Br adatoms bond ionically to the positively charged edge atoms of the nanoclusters. This conclusion is supported by the observed reduction in the NF for scattered $Na^+$, which shows that there is a change in the LEP above the edge atom sites, as well as the fact that the saturation coverages of Br are similar to those of CO provided in the literature.[50] The reduction in the LEP is presumably caused by the addition of downward pointing dipoles resulting from the adsorbed Br ions that oppose the upward pointing dipoles caused by the positively charged Au edge atoms. These edge sites are the same ones at which CO adsorbs[22, 59, 64], which indicates a dependence of the adsorption on the edge atom charge and the nanocluster size.[75] The catalytic activity of Au nanoclusters is known from previous studies to be reduced by halogen adsorption,[7, 73] so that the present results suggest that it is caused by site blocking. The results of this work further imply that the charge associated with the individual edge atoms in the clusters plays a critical role in the adsorption step of nanocatalytic reactions. Finally, the present results also suggest that a change in the NF of scattered alkali ions with cluster size may be a good indicator of a nanocluster's possible catalytic activity.

## ACKNOWLEDGEMENTS

This material is based upon work supported by the National Science Foundation under CHE - 1611563.

# REFERENCES

1. Lai, X.; Clair, T. P. S.; Valden, M.; Goodman, D. W. Scanning Tunneling Microscopy Studies of Metal Clusters Supported on TiO$_2$(110): Morphology and Electronic Structure. *Prog. Surf. Sci.* **1998,** *59*, 25-52.

2. Somorjai, G. A.; Li, Y. *Introduction to Surface Chemistry and Catalysis, Second Edition*. Wiley: Hoboken, NJ, 2010.

3. Huang, Z.; Geyer, N.; Werner, P.; de Boor, J.; Gösele, U. Metal-Assisted Chemical Etching of Silicon: A Review. *Adv. Mater.* **2011,** *23*, 285-308.

4. Winters, H. F.; Coburn, J. W. Surface Science Aspects of Etching Reactions. *Surf. Sci. Rep.* **1992,** *14*, 162-269.

5. Choy, K. L. Chemical Vapour Deposition of Coatings. *Prog. Mater Sci.* **2003,** *48*, 57-170.

6. Haruta, M.; Tsubota, S.; Kobayashi, T.; Kageyama, H.; Genet, M. J.; Delmon, B. Low-Temperature Oxidation of CO over Gold Supported on TiO$_2$ a-Fe$_2$O$_3$, and Co$_3$O$_4$. *J. Catal.* **1993,** *144*, 175-192.

7. Zhu, X.; Cheng, B.; Yu, J.; Ho, W. Halogen Poisoning Effect of Pt-TiO$_2$ for Formaldehyde Catalytic Oxidation Performance at Room Temperature. *Appl. Surf. Sci.* **2016,** *364*, 808-814.

8. Smith, D. P. Scattering of Low-Energy Noble Gas Ions from Metal Surfaces. *J. Appl. Phys.* **1967,** *38*, 340-347.

9. Weare, C. B.; Yarmoff, J. A. Resonant Neutralization of $^7$Li$^+$ Scattered from Alkali/Al(100) as a Probe of the Local Electrostatic Potential. *Surf. Sci.* **1996,** *348*, 359-369.

10. Kimmel, G. A.; Cooper, B. H. Dynamics of Resonant Charge Transfer in Low-Energy Alkali-Metal-Ion Scattering. *Phys. Rev. B* **1993,** *48*, 12164-12177.

11. Keller, C. A.; DiRubio, C. A.; Kimmel, G. A.; Cooper, B. H. Trajectory-Dependent Charge Exchange in Alkali Ion Scattering from a Clean Metal Surface. *Phys. Rev. Lett.* **1995,** *75*, 1654-1657.

12. Niehus, H.; Heiland, W.; Taglauer, E. Low-Energy Ion Scattering at Surfaces. *Surf. Sci. Rep.* **1993,** *17*, 213-303.

13. Los, J.; Geerlings, J. J. C. Charge Exchange in Atom-Surface Collisions. *Phys. Rep.* **1990,** *190*, 133-190.

14. Gauyacq, J. P.; Borisov, A. G. Charge Transfer in Atom-Surface Collisions: Effect of the Presence of Adsorbates on the Surface. *J. Phys.: Condens. Matter* **1998,** *10*, 6585-6619.

15. Weare, C. B.; Yarmoff, J. A. Resonant Neutralization of $^7$Li$^+$ Scattered from Cs/Al(100) as a Probe of the Local Electrostatic Potential. *J. Vac. Sci. Technol. A* **1995,** *13*, 1421-1425.

16. Canário, A. R.; Esaulov, V. A. Electron Transfer Processes on Ag and Au Clusters Supported on TiO$_2$(110) and Cluster Size Effects. *J. Chem. Phys.* **2006,** *124*, 224710.

17. Liu, G. F.; Sroubek, Z.; Yarmoff, J. A. Detection of Quantum Confined States in Au Nanoclusters by Alkali Ion Scattering. *Phys. Rev. Lett.* **2004,** *92*, 216801.

18. Arjad, A. B.; Yarmoff, J. A. Ion-Impact-Induced Strong Metal Surface Interaction in Pt/TiO$_2$ (110). *J. Phys. Chem. C* **2012,** *116*, 23377-23382.

19. Balaz, S.; Yarmoff, J. A. Low Energy Alkali Ion Scattering Investigation of Au Nanoclusters Grown on Silicon Oxide Surfaces. *Surf. Sci.* **2011,** *605*, 675-680.

20. Shen, J.; Jia, J.; Bobrov, K.; Guillemot, L.; Esaulov, V. A. Electron Transfer Processes on Au Nanoclusters Supported on Graphite. *Gold Bull.* **2013,** *46*, 343-347.

21. Shen, J.; Jia, J.; Bobrov, K.; Guillemot, L.; Esaulov, V. A. Electron Transfer Processes on Supported Au Nanoclusters and Nanowires and Substrate Effects. *J. Phys. Chem. C* **2015,** *119*, 15168-15176.

22. Salvo, C.; Karmakar, P.; Yarmoff, J. Inhomogeneous Charge Distribution Across Gold Nanoclusters Measured by Scattered Low Energy Alkali Ions. *Phys. Rev. B* **2018,** *98*, 035437.

23. Lemire, C.; Meyer, R.; Shaikhutdinov, S.; Freund, H. J. Do Quantum Size Effects Control CO Adsorption on Gold Nanoparticles? *Angew. Chem. Int. Ed.* **2004,** *43*, 118-121.

24. Tanaka, M.; Shudo, K.; Numata, M. Adsorption Site Preference of Br on Si(111)-7x7. *Phys. Rev. B* **2006,** *73*, 115326.

25. Takayanagi, K.; Tanishiro, Y.; Takahashi, M.; Takahashi, S. Structural Analysis of Si(111)-7×7 by UHV-Transmission Electron Diffraction and Microscopy. *J. Vac. Sci. Technol. A* **1985,** *3*, 1502-1506.

26. Yang, Y.; Yarmoff, J. A. Internal Charge Distribution of Iodine Adatoms on Silicon and Silicon Oxide Investigated with Alkali Ion Scattering. *Surf. Sci.* **2004,** *573*, 335-345.

27. Tromp, R.; Rubloff, G. W.; Balk, P.; LeGoues, F. K.; van Loenen, E. J. High-Temperature SiO$_2$ Decomposition at the SiO$_2$/Si Interface. *Phys. Rev. Lett.* **1985,** *55*, 2332-2335.

28. Zhang, L.; Cosandey, F.; Persaud, R.; Madey, T. E. Initial Growth and Morphology of Thin Au Films on TiO$_2$(110). *Surf. Sci.* **1999,** *439*, 73-85.

29. Varekamp, P. R.; Håkansson, M. C.; Kanski, J.; Shuh, D. K.; Björkqvist, M.; Gothelid, M.; Simpson, W. C.; Karlsson, U. O.; Yarmoff, J. A. Reaction of I$_2$ with the (001) Surfaces of GaAs, InAs, and InSb. I. Chemical Interaction with the Substrate. *Phys. Rev. B* **1996,** *54*, 2101-2113.

30. Spencer, N. D.; Goddard, P. J.; Davies, P. W.; Kitson, M.; Lambert, R. M. A Simple, Controllable Source for Dosing Molecular Halogens in UHV. *J. Vac. Sci. Technol. A* **1983,** *1*, 1554-1555.

31. Wang, W. K.; Simpson, W. C.; Yarmoff, J. A. Passivation versus Etching: Adsorption of $I_2$ on InAs(001). *Phys. Rev. Lett.* **1998,** *81*, 1465-1468.

32. Gao, R. S.; Gibner, P. S.; Newman, J. H.; Smith, K. A.; Stebbings, R. F. Absolute and Angular Efficiencies of a Microchannel-Plate Position-Sensitive Detector. *Rev. Sci. Instrum.* **1984,** *55*, 1756-1759.

33. Bedzyk, M. J.; Gibson, W. M.; Golovchenko, J. A. X-ray Standing Wave Analysis for Bromine Chemisorbed on Silicon. *J. Vac. Sci. Technol.* **1982,** *20*, 634-637.

34. Min, B. K.; Wallace, W. T.; Santra, A. K.; Goodman, D. W. Role of Defects in the Nucleation and Growth of Au Nanoclusters on $SiO_2$ Thin Films. *J. Phys. Chem. B* **2004,** *108*, 16339-16343.

35. Chang, C. S.; Knipping, U.; Tsong, I. S. T. Shadow Cones Formed by Target Atoms Bombarded by 1 to 3 keV $He^+$, $Li^+$, $Ne^+$ and $Na^+$ Ions. *Nucl. Instrum. Methods B* **1986,** *18*, 11-15.

36. Oen, O. S. Universal Shadow Cone Expressions for an Atom in an Ion Beam. *Surf. Sci.* **1983,** *131*, 407-411.

37. Rabalais, W. J. *Principles and Applications of Ion Scattering Spectrometry : Surface Chemical and Structural Analysis*. Wiley: New York, 2003.

38. Esaulov, V. A., Low Energy Ion Scattering and Recoiling Spectroscopy in Surface Science. In *Surface Science Techniques*, Braco, G.; Holst, B., Eds. Springer: Berlin, Germany, 2013; pp 423–460.

39. Jones, R. G. Halogen Adsorption on Solid Surfaces. *Prog. Surf. Sci.* **1988,** *27*, 25-160.

40. Wandlowski, T.; Wang, J. X.; Magnussen, O. M.; Ocko, B. M. Structural and Kinetic Aspects of Bromide Adsorption on Au(100). *J. Phys. Chem.* **1996,** *100*, 10277-10287.

41. Bertel, E.; Netzer, F. F. Adsorption of Bromine on the Reconstructed Au(100) Surface: LEED, Thermal Desorption and Work Function Measurements. *Surf. Sci.* **1980,** *97*, 409-424.

42. Zheleva, Z. V.; Dhanak, V. R.; Held, G. Experimental Structure Determination of the Chemisorbed Overlayers of Chlorine and Iodine on Au{111}. *Phys. Chem. Chem. Phys.* **2010,** *12*, 10754-10758.

43. Cochran, S. A.; Farrell, H. H. The Chemisorption of Iodine on Gold. *Surf. Sci.* **1980,** *95*, 359-366.

44. Parilis, E. S.; Kishinevsky, L. M.; Turaev, N. Y.; Baklitzky, B. E.; Umarov, F. F.; Verleger, V. K.; Nizhnaya, S. L.; Bitensky, I. S. *Atomic Collisions on Solid Surfaces*. North-Holland: Amsterdam, 1993.

45. Min, B. K.; Wallace, W. T.; Goodman, D. W. Support Effects on the Nucleation, Growth, and Morphology of Gold Nano-Clusters. *Surf. Sci.* **2006,** *600*, L7-L11.

46. Wang, Y.-G.; Cantu, D. C.; Lee, M.-S.; Li, J.; Glezakou, V.-A.; Rousseau, R. CO Oxidation on Au/TiO$_2$: Condition-Dependent Active Sites and Mechanistic Pathways. *J. Am. Chem. Soc.* **2016,** *138*, 10467-10476.

47. Roldan Cuenya, B.; Behafarid, F. Nanocatalysis: Size- and Shape-Dependent Chemisorption and Catalytic Reactivity. *Surf. Sci. Rep.* **2015,** *70*, 135-187.

48. Walsh, M. J.; Gai, P. L.; Boyes, E. E. On the Effect of Atomic Structure on the Deactivation of Catalytic Gold Nanoparticles. *J. Phys. Conf. Ser.* **2012,** *371*, 012048.

49. Arrii, S.; Morfin, F.; Renouprez, A. J.; Rousset, J. L. Oxidation of CO on Gold Supported Catalysts Prepared by Laser Vaporization: Direct Evidence of Support Contribution. *J. Am. Chem. Soc.* **2004,** *126*, 1199-1205.

50. Lemire, C.; Meyer, R.; Shaikhutdinov, S. K.; Freund, H. J. CO Adsorption on Oxide Supported Gold: From Small Clusters to Monolayer Islands and Three-Dimensional Nanoparticles. *Surf. Sci.* **2004,** *552*, 27-34.

51. Nordlander, P.; Tully, J. C. Energy Shifts and Broadening of Atomic Levels near Metal Surfaces. *Phys. Rev. B* **1990,** *42*, 5564-5578.

52. Gann, R. D.; Cao, J. X.; Wu, R. Q.; Wen, J.; Xu, Z.; Gu, G. D.; Yarmoff, J. A. Adsorption of Iodine and Potassium on Bi$_2$Sr$_2$CaCu$_2$O$_{8+\delta}$ Investigated by Low Energy Alkali Ion Scattering. *Phys. Rev. B* **2010,** *81*, 035418.

53. Borisov, A. G.; Teillet-Billy, D.; Gauyacq, J. P.; Winter, H.; Dierkes, G. Resonant Charge Transfer in Grazing Scattering of Alkali-Metal Ions from an Al(111) Surface. *Phys. Rev. B* **1996,** *54*, 17166-17174.

54. Weare, C. B.; German, K. A. H.; Yarmoff, J. A. Evidence for an Inhomogeneous-Homogeneous Transition in the Surface Local Electrostatic Potential of K-Covered Al(100). *Phys. Rev. B* **1995,** *52*, 2066-2069.

55. Zhang, Z.; Tang, W.; Neurock, M.; Yates Jr., J. T. Electric Charge of Single Au Atoms Adsorbed on TiO$_2$(110) and Associated Band Bending. *J. Phys. Chem. C* **2011,** *115*, 23848-23853.

56. Okazawa, T.; Fujiwara, M.; Nishimura, T.; Akita, T.; Kohyama, M.; Kido, Y. Growth Mode and Electronic Structure of Au Nano-Clusters on NiO(001) and TiO$_2$(110). *Surf. Sci.* **2006,** *600*, 1331-1338.

57. Visikovskiy, A.; Matsumoto, H.; Mitsuhara, K.; Nakada, T.; Akita, T.; Kido, Y. Electronic *d*-band Properties of Gold Nanoclusters Grown on Amorphous Carbon. *Phys. Rev. B* **2011,** *83*, 165428.

58. Zhang, C.; Michaelides, A.; King, D. A.; Jenkins, S. J. Positive Charge States and Possible Polymorphism of Gold Nanoclusters on Reduced Ceria. *J. Am. Chem. Soc.* **2010,** *132*, 2175-2182.

59. Vilhelmsen, L. B.; Hammer, B. Identification of the Catalytic Site at the Interface Perimeter of Au Clusters on Rutile TiO$_2$(110). *ACS Catalysis* **2014,** *4*, 1626-1631.

60. Zhong, W.; Zhang, D. Theoretical Study of Methanol Decomposition Mediated by Au$_3^+$, Au$_3$ and Au$_3^-$: Mechanism and Effect of Charge State of Gold on its Catalytic Activity. *Prog. React. Kinet. Mec.* **2013,** *38*, 86-94.

61. Sanchez, A.; Abbet, S.; Heiz, U.; Schneider, W. D.; Hakkinen, H.; Barnett, R. N.; Landman, U. When Gold is not Noble: Nanoscale Gold Catalysts. *J. Phys. Chem. A* **1999,** *103*, 9573-9578.

62. Yu, X.; Xu, L.; Zhang, W.; Jiang, Z.; Zhu, J.; Huang, W. Synchrotron-Radiation Photoemission Study of Growth and Stability of Au Clusters on Rutile TiO$_2$(110)-1x1. *Chin. J. Chem. Phys.* **2009,** *22*, 339-345.

63. Valden, M.; Lai, X.; Goodman, D. W. Onset of Catalytic Activity of Gold Clusters on Titania with the Appearance of Nonmetallic Properties. *Science* **1998,** *281*, 1647-1650.

64. Hong, S.; Rahman, T. S. Rationale for the Higher Reactivity of Interfacial Sites in Methanol Decomposition on Au$_{13}$/TiO$_2$(110). *J. Am. Chem. Soc.* **2013,** *135*, 7629-7635.

65. Choudhary, T. V.; Goodman, D. W. Oxidation Catalysis by Supported Gold Nano-Clusters. *Top. Catal.* **2002,** *21*, 25-34.

66. Flytzani-Stephanopoulos, M. Gold Atoms Stabilized on Various Supports Catalyze the Water-Gas Shift Reaction. *Acc. Chem. Res.* **2014,** *47*, 783-792.

67. Risse, T.; Shaikhutdinov, S.; Nilius, N.; Sterrer, M.; Freund, H. J. Gold Supported on Thin Oxide Films: From Single Atoms to Nanoparticles. *Acc. Chem. Res.* **2008,** *41*, 949-956.

68. Nilius, N.; Risse, T.; Shaikhutdinov, S.; Sterrer, M.; Freund, H.-J., Model Catalysts Based on Au Clusters and Nanoparticles. In *Gold Clusters, Colloids and Nanoparticles II*, Mingos, D. M. P., Ed. Springer: Berlin, 2014; Vol. 162, pp 91-138.

69. Bondzie, V. A.; Parker, S. C.; Campbell, C. T. The Kinetics of CO Oxidation by Adsorbed Oxygen on Well-Defined Gold Particles on TiO$_2$(110). *Catal. Lett.* **1999,** *63*, 143–151.

70. Shaikhutdinov, S. K.; Meyer, R.; Naschitzki, M.; Bäumer, M.; Freund, H.-J. Size and Support Effects for CO Adsorption on Gold Model Catalysts. *Catal. Lett.* **2003,** *86*, 211-219.

71. Mørk, P. C.; Norgård, D. Nickel-Catalyzed Hydrogenation: A Study of the Poisoning Effect of Halogen-Containing Compounds. *J. Am. Oil Chem. Soc.* **1976,** *53*, 506-510.

72. Bartholomew, C. H. Mechanisms of Catalyst Deactivation. *Appl. Catal. A* **2001,** *212*, 17-60.

73. Gracia, F. J.; Miller, J. T.; Kropf, A. J.; Wolf, E. E. Kinetics, FTIR, and Controlled Atmosphere EXAFS Study of the Effect of Chlorine on Pt-Supported Catalysts during Oxidation Reactions. *J. Catal.* **2002,** *209*, 341-354.

74. Pettersson, L. G. M.; Bagus, P. S. Adsorbate Ionicity and Surface-Dipole-Moment Changes: Cluster-Model Studies of Cl/Cu(100) and F/Cu(100). *Phys. Rev. Lett.* **1986,** *56*, 500-503.

75. Min, B. K.; Deng, X. Y.; Li, X. Y.; Friend, C. M.; Alemozafar, A. R. Tuning Reactivity and Selectivity for Olefin Oxidation through Local O Bonding on Au. *ChemCatChem* **2009,** *1*, 116-121.